\documentclass[iop]{emulateapj}
\shorttitle{A new candidate \gray\ pulsar}
\shortauthors{Strader \etal~}
\def\etal{{et al.}}
\def\kms{\,km~s$^{-1}$}

\def\arcsec{\char'175 }

\def\hub{\ifmmode H_\circ\else H$_\circ$\fi}
\usepackage{subfigure}
\usepackage{amsmath}
\usepackage{graphics}
\usepackage{hyperref}
\usepackage{breakurl} 

\def\ltsima{$\; \buildrel < \over \sim \;$}
\def\simlt{\lower.5ex\hbox{\ltsima}} 
\def\gtsima{$\; \buildrel > \over \sim \;$}
\def\simgt{\lower.5ex\hbox{\gtsima}} 
\def\arcsec{\hbox{$^{\prime\prime}$}}
\def\deg{\hbox{$^\circ$}}

\def\gray{$\gamma$-ray}

\def\Fermi{\textit{Fermi}}

\def\zJ0523{1FGL~J0523.5--2529}

\begin{document}

\title{1FGL~J0523.5--2529: A New Probable Gamma-ray Pulsar Binary}
\author{
Jay Strader\altaffilmark{1},
Laura Chomiuk\altaffilmark{1},
Eda Sonbas\altaffilmark{2},
Kirill Sokolovsky\altaffilmark{3,4},
David J.~Sand\altaffilmark{5},
Alexander S.~Moskvitin\altaffilmark{6}
C.~C.~Cheung\altaffilmark{7}}

\altaffiltext{1}{Department of Physics and Astronomy, Michigan State University, East Lansing, MI 48824, USA}
\altaffiltext{2}{Department of Physics, University of Adiyaman, 02040 Adiyaman, Turkey}
\altaffiltext{3}{Astro Space Center of the Lebedev Physical Institute, 117810 Moscow, Russia}
\altaffiltext{4}{Sternberg Astronomical Institute of the Moscow State University, 119992 Moscow, Russia}
\altaffiltext{5}{Department of Physics, Texas Tech University, Box 41051, Lubbock, TX 79409}
\altaffiltext{6}{Special Astrophysical Observatory, Russian Academy of Sciences, Nizhnii Arkhyz, Karachaevo-Cherkesskaya Republic, 369167 Russia}
\altaffiltext{7}{Space Science Division, Naval Research Laboratory, Washington, DC 20375, USA}

\begin{abstract}

We report optical photometric and SOAR spectroscopic observations of an X-ray source found within the localization error of the \Fermi-LAT unidentified \gray\ source \zJ0523. The optical data show periodic flux modulation and radial velocity variations indicative of a binary with a 16.5-hr period. The data suggest a massive non-degenerate secondary ($\ga 0.8 M_{\odot}$), and we argue the source is likely a pulsar binary. The radial velocities have good phase coverage and show evidence for a measurable eccentricity ($e=0.04$). There is no clear sign of irradiation of the secondary in either photometry or spectroscopy. The spatial location out of the Galactic plane and \gray\ luminosity of the source are more consistent with classification as a recycled millisecond pulsar than as a young pulsar. Future radio timing observations can confirm the identity of the primary and further characterize this interesting system.

\end{abstract}


\section{Introduction\label{section-intro}}

The \emph{Fermi} \gray\ Space Telescope has changed our understanding of the high-energy sky at GeV energies. While many $\gamma$-ray sources are distant active galactic nuclei, a large number of Galactic sources have been classified, including pulsars, supernova remnants, pulsar wind nebulae and novae (Nolan et al. 2012). However, there is still a substantial population of unidentified sources in the \emph{Fermi} Large Area Telescope (LAT) catalog (Atwood et al.\ 2009). Further study of these unidentified sources is motivated by a desire to enlarge the sample of still-rare classes of $\gamma$-ray sources (like pulsar binaries) and to ensure that we have fully mapped $\gamma$-ray discovery space.

Follow-up of high-latitude \emph{Fermi} sources has yielded critical insights into the formation and evolution of millisecond pulsars. While millisecond pulsars have long been thought to be spun up and ``recycled" through accretion from a binary companion, only a few millisecond pulsars in the Galactic field\footnote{A population of these systems is known in globular clusters, but it has proven difficult to study these in detail; see the catalog at \burl{http://www.naic.edu/~pfreire/GCpsr.html}} were confirmed to have non-degenerate binary companions before the launch of \emph{Fermi} (Fruchter et al.\ 1990, Stappers et al.\ 1996, Burgay et al.\ 2006; see also the review of Lorimer 2005). \emph{Fermi} has proven an effective tool for growing this sample, as millisecond pulsars are efficiently detected in GeV $\gamma$-rays (e.g., Abdo et al.\ 2013). Further, it is possible that in some binary systems the pulsar $\gamma$-ray emission can be enhanced as the pulsar wind interacts with the stellar wind of its binary companion and accelerates particles to relativistic speeds (Bednarek \& Sitarek 2013).

Owing to follow-up of \emph{Fermi} sources, the sample of field eclipsing millisecond pulsar binaries has grown more than six-fold over the past four years (Roberts 2013). Eclipsing pulsar binaries are often divided into ``black widow" systems, in which the companion is non-degenerate but low mass ($\lesssim$0.05 M$_{\odot}$) due to ablation by the pulsar wind, and ``redback" systems with more massive non-degenerate companions ($\gtrsim 0.15 M_{\odot}$). Some redbacks show continuing accretion onto the pulsar, hinting that they may be transition systems between accreting low-mass X-ray binaries and non-accreting millisecond pulsars (Wang et al.\ 2009). However, despite this recent progress, the sample of redback systems remains small, hampering our ability to explore the parameter space of millisecond pulsar progenitors.

The follow-up process for \emph{Fermi} unidentified sources usually spans the electromagnetic spectrum. For example, Cheung et al.~(2012) used \emph{Chandra} to obtain X-ray images over the error ellipses of several Fermi-LAT sources out of the Galactic plane. With the refined X-ray positions in hand, variable optical counterparts were found for two sources that were then proposed to be elusive ``radio-quiet" millisecond pulsars (Kong et al. 2012; Romani \& Shaw 2011; Romani 2012; Kataoka et al. 2012). Both systems were subsequently confirmed as pulsars at radio wavelengths (Pletsch et al. 2012; Belfiore 2013; Ray et al. 2013); time-variable mass loss may sometimes obscure the radio emission.

Using similar techniques, here we present the discovery of a likely new pulsar binary associated with the \Fermi-LAT unidentified \gray\ source that appeared in the 1FGL catalog as  \zJ0523 (Abdo et al.~2010). Based on two years of observations (Nolan et al.~2012), this source was re-catalogued as 2FGL~J0523.3$-$2530. Its 0.1--100 GeV emission is characterized as a single power-law with photon index = $2.12 \pm 0.07$ with no significant variability\footnote{See the 2FGL LAT 1-month binned light curve: \burl{http://heasarc.gsfc.nasa.gov/FTP/fermi/data/lat/catalogs/source/lightcurves/2FGL_J0523d3m2530_lc.png}}. No radio emission from a pulsar has yet been detected in this system: Guillemot \etal~(2012) and Petrov \etal~(2013) present non-detections to $5\sigma$ flux densities of $\sim 60$--70 $\mu$Jy (1.4 GHz) and $\sim 1$ mJy (5.5 and 9 GHz), respectively. 
 
Nonetheless, on the basis of the data presented in this paper, we classify  \zJ0523 as a probable binary pulsar with an unusually massive ($\ga 0.8 M_{\odot}$) secondary. 

\section{Observations\label{sec-observations}}

\subsection{X-ray Counterpart}

1FGL~J0523.5$-$2529 was observed with the Swift/XRT on 2009 November 12 with an exposure time of 4.8 ksec. The image yielded a single candidate X-ray source counterpart within the 1FGL catalog error circle, which had a 0.3--10 keV count rate $3.08\pm0.94$ cts ksec$^{-1}$ ($3.3 \sigma$) and is located at a J2000 sexigesimal position of (R.A, Dec.) = (05:23:17.18, --25:27:37.4) \citep{tak13}. The X-ray source is offset by 0.043\deg\ from the subsequent 2FGL catalog position (RA = 80.828\deg, Decl. = --25.503\deg), which has a 95$\%$ confidence error radius $\sim 0.06$\deg. This source is the focus of this paper. We note the 90\% error radius for the X-ray position is 6.2\arcsec\ and may be the counterpart of a ROSAT-detected source, 1WGA~J0523.2--2527 \citep{whi00}. The inferred 0.3--10 keV X-ray luminosity from the Swift data is $\sim 1.6 \times 10^{31} (d/1.1\, \textrm{kpc})^2$ erg s$^{-1}$ (see \S 3.3 for discussion of the source distance).

\subsection{Optical Photometry}

Using the USNO B1.0 catalog \citep{mon03}, we found an optical counterpart (USNO-B1.0~0645-0058532) for the X-ray source with magnitudes $B2=17.38$, $R2=15.92$, $I=15.94$ mag, at (R.A, Dec.) = (05:23:16.925, --25:27:36.92), offset from the X-ray position by only 3.5\arcsec. There is a near-infrared match (0.03\arcsec\ offset) to the USNO-B1.0 source, 2MASS~J05231692-2527369, with $J=14.89 \pm 0.04$, $H=14.44\pm 0.04$, and $K=14.29 \pm 0.07$ mag \citep{skr06}. 

We obtained archived photometric observations of the USNO B1.0 source from the Catalina Sky Survey \citep[CSS;][]{dra09} and its sister Siding Spring Survey (SSS). The CSS covers the sky with unfiltered images in sequences of four 30-sec exposures typically reaching $V$-equivalent magnitudes of  $\sim 19$--20 mag. We found the source catalogued in both the CSS and SSS databases as J052316.9--252737, with 233 data points over a long time range from 2005 August 17 to 2013 April 12. The average magnitude was $V \sim 16.5$, with typical uncertainties per epoch of  0.07 to 0.13 mag.

We used {\tt Period04} (Lenz \& Breger 2005) to determine a best-fit optical photometric period of $0.688139\pm0.000085$ d. A period about half this (0.34406 d) gives a comparably good fit (with only one maximum per orbit), but is an alias: the longer period matches that obtained independently from spectroscopy (0.688134 d; \S 3.1). Figure 1 shows the CSS/SSS photometry phased with this spectroscopic period, which is consistent with the photometric period to within 0.4 sec. We use the convention that superior conjunction (when the secondary is ``full", i.e., behind the primary along the line of sight) is at $\phi = 0.5$.

\begin{figure}[ht]
\includegraphics[width=3.3in]{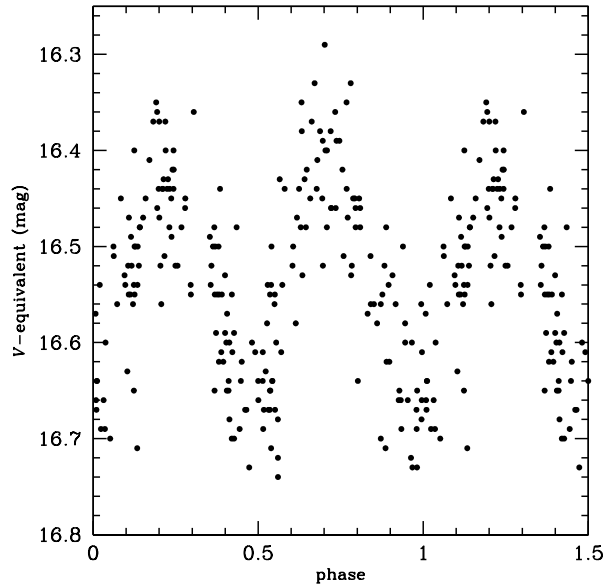}
\caption{Phased CSS/SSS unfiltered photometry of J0523.5--2529 using the spectroscopic period 0.688134 d and phase. Only points with uncertainties $< 0.2$ mag are plotted.}
\end{figure}

\subsection{Optical Spectroscopy}

Following the identification of a candidate optical photometric period, we initiated spectroscopic follow-up with the Goodman High-Throughput Spectrograph (Clemens \etal~2004) on the SOAR 4.1-m telescope. On the first night we took two spectra with a low-resolution grating and 1.03\arcsec\ slit (400 l mm$^{-1}$; resolution 5.7 \AA). One of the spectra is shown in Figure 2. All subsequent observations used a 2100 l mm$^{-1}$ grating and a 1.03\arcsec\ slit, giving a resolution of 0.9 \AA\ (Figure 2). For most observations the wavelength range was $\sim 4960$--5600 \AA. 

Individual spectra were taken on six nights over the period 2013 October 1 to 2014 January 10, with typical exposure times of 5 min. A large radial velocity change was observed between the first and second observing nights, after which more frequent spectra were obtained. FeAr arcs were taken for wavelength calibration every 30 min throughout the night.

The spectra were reduced in the standard manner, with optimal extraction and wavelength calibration using arcs. To help correct for the effects of flexure, we used the bright 5577.34 \AA\ sky line to make small zeropoint shifts to the wavelength solution for each spectrum.

Visually, the optical spectra appear consistent with a late G or early K spectral type.  A radial velocity was derived for each spectrum through cross-correlation with a set of bright stars of similar spectral type taken with the same instrumental setup. Examples of both low- and high-resolution spectra are plotted in Figure 2. There is no evidence for more than one set of spectral lines in any of the spectra.

\begin{figure}[ht]
\includegraphics[width=3.3in]{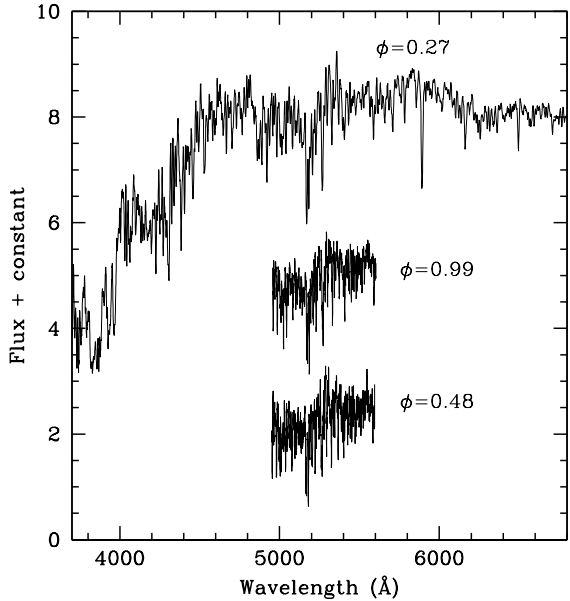}
\caption{Sample low-resolution and high-resolution optical spectra, smoothed with a 5-pixel boxcar for display. The corresponding phases are listed. The high-resolution data are plotted near inferior and superior conjunction. No clear differences in spectral shape or individual spectral lines are seen, suggesting irradiation may not be important in the system.}
\end{figure}

\section{Results}

\subsection{Orbital Fitting}

After converting the mid-exposure observing times to Barycentric Julian Date (BJD) on the Barycentric Dynamical Time (TDB) system (Eastman \etal~2010), we did a Keplerian fit to the radial velocity measurements using the IDL package {\tt BOOTTRAN} (Wang et al.~2012), with uncertainties calculated via bootstrap. Figure 3 shows the phased radial velocities with the best fit overplotted. Formally this fit is excellent ($\chi^2$ = 106/90 d.o.f.), with an r.m.s. residual of 7.0 \kms. The best-fit parameters and uncertainties are period $P = 0.688134\pm0.000028$ d, secondary semi-amplitude $K_2=190.3\pm1.1$ \kms, eccentricity $e = 0.040\pm0.006$, longitude of periastron $\omega = 214\pm10 \deg$, and a systemic velocity of $57.1\pm0.9$ \kms. In lieu of the time of periastron (which is poorly constrained due to the low $e$), we give the BJD at the superior conjunction ($\phi = 0.5$; when the secondary is behind the primary) preceding the first radial velocity measurement, which is $T_{0.5} = 2456577.64636\pm0.0037$ d. As mentioned in \S 2.2, the best-fit radial velocity period is consistent with the photometric period to within 0.4 sec. The mass function of the secondary is $f(M) = 0.49\pm0.01 M_{\odot}$. 

The evidence for the presence of at least apparent eccentricity is strong: if we fix $e=0$ the resulting $\chi^2$ = 161/91 d.o.f., substantially worse. In irradiated binaries a ``false" eccentricity is sometimes observed in which the center-of-mass of the system is offset from the center-of-light, due to the change in the effective temperature of the visible region of the secondary during the orbit. However, we do not see evidence for irradiation: there 
is no significant change in spectral type as a function of phase nor are there different mean magnitudes at inferior and superior conjunction (see Figures 1 and 2).

Most binaries that emit $\gamma$-rays have neutron star primaries (Dubus 2013), so this is a natural starting point for the interpretation. Irradiation might be expected for a pulsar primary: the total integrated \gray\ luminosity (0.1--100 GeV) in 2FGL, assuming no beaming, is $\sim 3.1\times10^{33} (d/1.1\, \textrm{kpc})^2$ erg s$^{-1}$ (Nolan et al.~2012; see \S 3.3 for distance estimate), which is typical of millisecond pulsars detected with \emph{Fermi} (Abdo et al.~2013). The implied spindown luminosity is $\ga 10^{34}$ erg s$^{-1}$, comparable to known redbacks (Roberts 2013). One possible explanation for the apparent lack of irradiation is if the pulsar radiation is beamed so that it misses the secondary, or hits only a small portion of its surface. The high mass of the secondary compared to most redbacks may also help in minimizing the effects of irradiation. 

In any case, for now we consider it possible that irradiation is unimportant in the system and thus that the eccentricity is in fact greater than zero. Nearly all field millisecond pulsars and redbacks have negligible eccentricity (e.g., Phinney \& Kulkarni 1994, Crawford et al.~2013), suggesting that \zJ0523 could be a partially-recycled system not yet circularized, or perhaps that it is in a triple with the eccentricity induced via the Kozai mechanism (e.g., Champion et al.~2008). However, we cannot rule out other explanations for the apparent eccentricity, such as a small systematic error in a subset of the measured radial velocities.

The 16.5 hr period and 190 \kms\ semi-amplitude, combined with the apparent late-G/early-K spectral type of the secondary, are consistent with a neutron star primary with a non-degenerate main sequence companion. Single solar metallicity stars of this spectral type have masses of $\sim 0.8$--$1.0 M_{\odot}$ (the masses would be lower if the assumed metallicity were reduced, though the spectra in Figure 2 suggest a relatively metal-rich star). While the implied companion is unusually massive compared to most known redbacks (Roberts 2013), as we show below, the full set of spectroscopic data provide additional support for a secondary mass in this range. 

\subsection{Rotational Velocity}

Given the short period of the binary, the secondary star is expected to be spun up and tidally locked to the primary (the synchronization timescale is $\la 10^4$ yr for the mass ratio found below; Zahn 1977). The rotational velocity of the secondary, when combined with the orbit semi-amplitude, can be used to constrain the mass ratio of the binary.

The resolution of our spectra is high enough ($R\sim 6000$) that we can estimate the projected rotational velocity $V_{rot}$ sin $i$. We do this in the usual manner for moderate resolution data (e.g., Rhode \etal~2001) by  convolving standard stars with a set of kernels reflecting a range of $V_{rot}$ sin $i$ values and that include limb darkening. We then fit relations between the set of input values of $V_{rot}$ sin $i$  and the resulting full-width half-maximum (FWHM) values for the resulting spectra. We then use the FWHM values derived from cross-correlating the spectra of \zJ0523 with the standard stars to estimate the rotational velocity.

The value obtained in this manner is $V_{rot}$ sin $i = 102\pm6$ \kms. The quoted uncertainty in this value is due only to random errors. Higher-resolution spectra, combined with a more realistic stellar model, could improve this measurement in the future.

\begin{figure}[ht]
\includegraphics[width=3.3in]{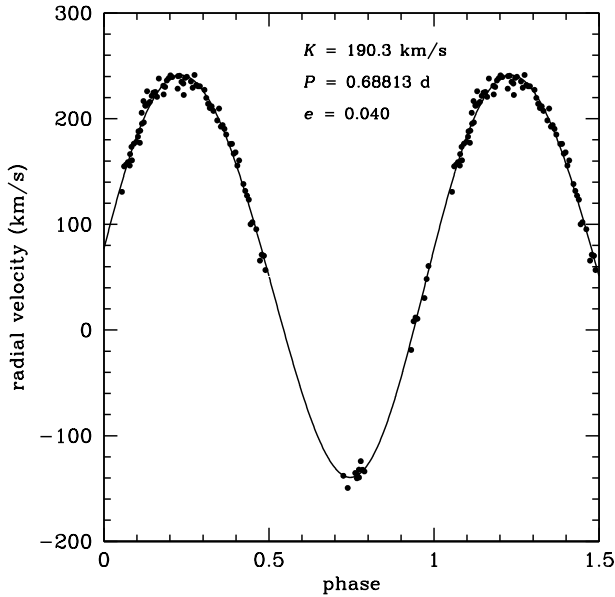}
\caption{Phased radial velocities for the optical counterpart to 1FGL J0523.5--2529, with the listed Keplerian fit overplotted.}
\end{figure}

\subsection{Other Binary Parameters}

The measured rotational velocity and semi-amplitude of the secondary $K_2$ immediately give the mass ratio $q = M_2/M_1$ of the binary via the standard formula (Casares 2001):

\begin{equation}
 V_{rot} \, \textrm{sin} \, i = 0.46\, K_2\, q^{1/3}\, (1+q)^{2/3}
\end{equation}

\noindent
This implies $q = 0.61\pm0.06$. Most binaries that emit $\gamma$-rays have neutron star primaries (Dubus 2013). If we assume the primary has a normal neutron star mass of 1.4 $M_{\odot}$, then the secondary mass is $M_2 = 0.85\pm0.08 M_{\odot}$, consistent with that inferred simply from its spectral type. If the primary mass were as high as 2.0 $M_{\odot}$, the inferred secondary mass would be in the range 1.1--1.3 $M_{\odot}$. The corresponding inclinations range from 75\deg\ to 59\deg\ for 1.4 to 2.0 $M_{\odot}$ primaries.

With the current data we cannot conclusively exclude other compact objects as primaries, including white dwarfs and black holes. However, these classifications appear to be very unlikely: the only white dwarfs known to be \gray\ sources are novae in outburst (Dubus 2013), and the spectral type of the secondary and the inferred binary mass ratio are inconsistent with even a relatively low-mass black hole.

We also considered the possibility that there is a chance superposition of the X-ray source with the \gray\ source, in which case the system could be interpreted as an active binary such as a W UMa (contact binary). Since we observe a single-lined binary, the visible star would need to be the primary, rather than the secondary as assumed above. However, given this assumption, there is no plausible set of binary parameters consistent with the observed semi-amplitude and period. Thus we do not believe this is a viable explanation for this source.

In general, it is possible to obtain additional constraints on the binary properties---including the inclination and Roche lobe filling factor---using photometry. While the CSS and SSS photometric data are nominally reported in $V$, the images are unfiltered, so there are significant color terms that relate these magnitudes to standard $V$, and they are ill-suited for detailed model fitting. We can make the qualitative statement that the minima at superior and inferior conjunction of the ellipsoidal variations do not show significant differences, suggesting that the secondary may not be close to filling its Roche lobe.

If we assume that the secondary is indeed an early K dwarf that is relatively unaffected by irradiation, we can use the Padova isochrones (Bressan \etal~2012) to estimate the distance to the binary. Using the 2MASS $K$-band photometry to minimize the effects of extinction, we infer a distance of $\sim 1.1\pm0.3$ kpc, a value typical for {\it Fermi} \gray\ binary pulsars (e.g., Abdo \etal 2013, Gentile \etal~2014). However, we emphasize this value depends strongly on the assumed mass.

\section{Discussion and Summary \label{sec-discussion}}

The most straightforward model consistent with all available data is that the \gray\ source \zJ0523 is a 1.4 $M_{\odot}$ pulsar with a relatively massive, $\sim 0.8$--0.9 $M_{\odot}$ main sequence secondary, though systems with more massive components are not ruled out. However, the physical properties of the pulsar are uncertain.

As discussed in the Introduction, many of the new pulsar binaries discovered via follow-up of \emph{Fermi} sources have been classified as redbacks: eclipsing millisecond pulsars with $\ga0.15M_{\odot}$ secondaries being ablated by the pulsar wind. Besides the circumstantial association of redbacks with binary pulsars discovered via \emph{Fermi} follow-up, the principle evidence that \zJ0523 is a redback is that it has a main sequence secondary, and that no radio emission was detected in two searches (Guillemot \etal~2012; Petrov \etal~2013), consistent with the temporary obscuration of the pulsar radio emission as observed in some other redbacks (e.g., Ray et al. 2013). However, this is indirect evidence for ablation, and as discussed in \S 3.1 there is no direct evidence for irradiation. In addition, the inferred large mass of the secondary and the non-zero eccentricity are both unusual compared to confirmed redbacks in the field (e.g, Roberts 2013; Crawford et al.~2013).

An alternative possibility is that the primary is a young pulsar rather than a recycled pulsar, which could explain both the eccentricity and high secondary mass. However, other aspects of the system are less consistent with this hypothesis. The \gray\ luminosity of $\sim 3.1\times10^{33} (d/1.1\, \textrm{kpc})^2$ erg s$^{-1}$ is typical of millisecond pulsars but low compared to that of young pulsars (although not exceptionally so; Abdo et al.~2013). Nearly all young pulsars in the field detected with \emph{Fermi} are located close to the Galactic plane, while \zJ0523 is at approximate Galactic coordinates ($l = 228.2\deg$, $b=-29.8\deg$), corresponding to a location $\sim 550$ pc below the plane for the assumed distance of 1.1 kpc. In addition, the small sample of young field pulsars in binaries with main sequence stars all have B-type or earlier companions and long periods (Dubus 2013), dissimilar to \zJ0523.

Additional observations are necessary to better understand this unusual source. In particular, timing observations to discover the suspected pulsar should be a high priority.

Only a handful of binary pulsars have been found through optical follow-up of unidentified Fermi-LAT sources. This work shows that it is likely more such systems remain to be discovered.

\acknowledgments

We are grateful for the comments of an anonymous referee. These comments significantly improved the paper.

Based on observations obtained at the Southern Astrophysical Research (SOAR) telescope, which is a joint project of the Minist\'{e}rio da Ci\^{e}ncia, Tecnologia, e Inova\c{c}\~{a}o (MCTI) da Rep\'{u}blica Federativa do Brasil, the U.S. National Optical Astronomy Observatory (NOAO), the University of North Carolina at Chapel Hill (UNC), and Michigan State University (MSU).

The CSS survey is funded by the National Aeronautics and Space Administration under Grant No.~NNG05GF22G issued through the Science Mission Directorate Near-Earth Objects Observations Program.  The CRTS survey is supported by the U.S.~National Science Foundation under grants AST-0909182.

Work by C.C.C. at NRL is supported in part by NASA DPR S-15633-Y.

{}

\end{document}